\documentstyle[preprint,aps]{revtex}

\begin{document}
\title{On the spectroscopic structure of two interacting electrons in a quantum dot }
\author{Omar Mustafa}
\address{Department of Physics, Eastern Mediterranean University\\
G. Magusa, North Cyprus, Mersin 10 - Turkey\\
email: omar.mustafa@emu.edu.tr\\
}
\maketitle

\begin{abstract}
{\small The shifted 1/N expansion technique used by El-Said (Phys. Rev. {\bf %
B 61}, 13026 (2000)), to study the relative Hamiltonian of two interacting
electrons confined in a quantum dot, is investigated. El-Said's results from
SLNT are revised and results from an alternative method are also reported.
The distinctive role of the central spike term, }$(m^{2}-1/4)/q^{2}$, 
{\small in determining spectral properties of the above problem is shown,
moreover.}
\end{abstract}

\newpage

In his paper (Phys. Rev. {\bf B 61}, 13026 (2000)) El-Said [1] used the
shifted 1/N expansion technique to study spectroscopic structure of the two
- dimensional ({\em flat land}) motion of two electrons ( with effective
mass $m^{\ast }$), confined by a harmonic potential of characteristic length 
$l_{o}=\sqrt{\hbar /(m^{\ast }\omega _{o})}$ in the xy-plane.

The Hamiltonian of this problem is known to decouple into two quasi -
particle Hamiltonians: the center of mass and relative motion ones [1-5].
The center of mass Hamiltonian is exactly soluble and the relative motion
one \newline
\begin{equation}
\left[ -\frac{d^{2}}{dq^{2}}+\frac{(m^{2}-1/4)}{q^{2}}+\frac{\lambda }{q}%
+q^{2}\right] U_{k,m}(q)=\Xi _{k,m}U_{k,m}(q),
\end{equation}
\newline
( with $q=r/(\sqrt{2}l_{o})$, $\Xi _{k,m}=2E_{k,m}/(\hbar \omega _{o})$, $%
\lambda =\sqrt{2}l_{o}/a^{\ast },$ and $k$ is the radial quantum number) is
known to belong to non-exactly soluble Hamiltonians. Therefore, one has to
resort to approximation methods like SLNT, used by El-Said [1].

In fact, the up-normal performance of SLNT, reported in table I of [1],
attracted our attention and inspired the current report. It is well known
that because of the additive complexity in handling large - order
corrections of the standard Rayleigh - Schr\"{o}dinger perturbation theory,
only low - order corrections ( up to the third - order) have been reported
for SLNT [6-9]. Eventually, the results of SLNT are not as accurate and
reliable as sought after ( documented by Mustafa and Odeh [10], Fernandez et
al. [11], Maluendez et al. [12], and others [13]).

The actual results of SLNT ( following SLNT procedure in [1,6-9]) are listed
in table I, along with those reported by El-Said [1] and the exact numerical
integration ones [2]. Obviously, sever deviations ( underlined) from the
exact ones occur for $k\geq 1$ results. Moreover, a level - ordering change
is clearly manifested for the $(1,3)$ and $(2,0)$ states. Evidently, the
results from SLNT [1] can not address the sensitive spectral properties of
the attendant problem, especially when levels - ordering, energy -
crossings, spin- oscillations, magnetic - fingerprints [5], $\cdots $etc,
are in point. Nevertheless, results from perturbation theory are limited to
the case where $\lambda \ll 1$ [14,15] and those from WKB treatments [2]
lead to dubious accuracies (in connection with level ordering and energy
crossings).

In numerous methodical predecessors of a subset of papers [5,13,16-18], an
alternative possibility has been sought in the power-law asymptotic
expansions using some {\em small} parameter. It has been noticed that the
presence of the central spike, e.g. $(m^{2}-1/4)/q^{2}$ in (1), just {\em %
copies} the effect of the centrifugal and/or centripetal force and
immediately inspires the use of {\em small} shifted inverse angular momentum
quantum number (PSLET). An exhaustive description of the necessary formulae
of PSLET accompanied by the persuasive verifications of their numerical
usefulness ( by immediate comparisons of its results with available {\em %
brute force} numerical data) could be found in ref.s [5,13,16-18].

Following PSLET procedure we compare our results, in table II, with the
exact numerical ones ( obtained by direct numerical integrations, DNI) [2]
for $\lambda =1$ and $\lambda =10$. To avoid exhaustive numbers of tables we
do not list Garcia-Castelan et. al's results [2] from WKB, WKB
single-parabola (WKB-SP), and WKB double-parabola (WKB-DP). In contrast with
the WKB, WKB-SP, WKB-DP [2] and SLNT [1] results, the comparison between
PSLET and DNI results implies excellent agreement.

In order to make remediable analysis on the effect of $\lambda $, hence of
the characteristic length $l_{o}$ ($\lambda \sim \l _{o}$), we list ( in
tables III-V) PSLET results for $k=0,1,2$ and $\lambda =0,1,2,4,6,8,10,12$
at different values of $|m|$. They are also plotted in figure I.

Figure I (along with tables III-V) shows that the degeneracies associated
with the harmonic oscillator confinement at $\lambda =0$ are only partially
lifted as $\lambda $ increases from zero ( of course, such degeneracies
would completely be lifted when a magnetic field is applied perpendicular to
the plane of the dot). It also shows that the equidistance form of the
energy levels at $\lambda =0$ changes in the following manners; (i) for a
given $k$, the spacing between two successive $|m|$ states decreases as $%
\lambda $ increases, and increases as $|m|$ increases for a given $\lambda $%
, whilst (ii) for a given $\lambda ,$ the spacing increases as the nodal
quantum number $k$ increases. One should nevertheless notice that (iii)
s-states (with $m=0$) shift up more rapidly than states with $|m|\geq 1$,
and for $|m|\geq 1$ states with lower $|m|$ shift up faster than states with
higher $|m|$ as $\lambda $ increases from zero.

The above mentioned features (i)-(iii), in fact, build up the sought after
scenario for the change in level ordering, that manifests energy crossings
and spin-singlet ($S_{z}=0$) spin-triplet ($S_{z}=1$) oscillations, and
inspires the vital role of the central spike term in (1). More specifically,
the {\em twofold} nature of the central spike term in the effective potential
\begin{equation}
V_{eff}(q)=\frac{m^{2}-1/4}{q^{2}}+q^{2}+\frac{\lambda }{q}
\end{equation}
\newline
explains the energy crossings as follows; (a) for $m=0$ it represents an
attractive core that strengthens the confinement $q^{2}$, whereas (b) for $%
|m|\geq 1$ it represents a repulsive core which renders, along with the
Coulomb repulsion, the potential less potent. This is why, for a given $k$,
the energy of a lower $|m|$ state increases much faster ( more rapidly for $%
m=0$) than that of a higher $|m|$, as $\lambda $ increases, and catches up
with it ( hence energy crossings and singlet-triplet spin oscillations
occur, or, at most, energy levels clustering is manifested). On the physical
sides, the two electrons are farther apart for higher $|m|$. Moreover, for a
given $k$ energy crossings are not feasible between the corresponding states
with different $|m|$. Whereas, states with a given $k$ and $|m|$ cross with
states at lower $k$ and higher $|m|$. Therefore, the lowest three states
(0,0), (0,1), and (0,2) never cross any other state ( i.e., they can never
be depressed into a lower $k$-state).

The effect of correlation, between two interacting electrons in a harmonic
QD, is therefore clear in the full energy spectrum for $\lambda >0$ with all
($k$,$|m|$)-states for the relative motion as shown in figure I and
documented in tables II-V. However, it should be noted that the level
ordering reported by Garcia-Castelan et. al [2] is now changed, namely for
the (0,4) and (1,1) states. Moreover, the (1,4) and (2,1) states seem to
change order as $\lambda $ increases from 12.

To sum up, we have used SLNT [1,6-8] and revised the numerical results
reported by El-Said [1], on the correlation energies for two interacting
electrons in a parabolic QD. As an alternative, we have used a
pseudo-perturbation recipe PSLET [5,13,16-18] and proved PSLET persuasive
numerical reliability in comparison with direct numerical integration method
( in table II) for $\lambda $=1, and 10. In contrast with perturbation
theory results, which are limited to $\lambda <<1$ [14,15], we have obtained
the correlation energies for $\lambda =2,4,6,8,12$ and $k=0,1,2$. We have
documented ( through figure I) that the level ordering reported by
Garcia-Castelan et. al [2] is not absolute but bound to change as $\lambda $
increases from zero. Herein, one should report that results from WKB
treatments [2] lead to poor accuracy for states with $k=0$ and/or $m=0$.
Moreover, the {\em twofold} effect of the central spike term, in the
effective two-dimensional potential (2), is now clarified to inherit a major
responsibility for energy crossings.

\newpage 
\begin{table}[tbp]
\caption{Comparison of PSLET energies (in $\hbar\protect\omega_o/2$ units)
and the exact ones from direct numerical integration [2] for $\protect%
\lambda $=1 and 10.}
\begin{center}
\vspace{1cm} 
\begin{tabular}{|cccccc|}
\hline\hline
& $\lambda$=1 &  &  & $\lambda$=10 &  \\ \hline
$(k,|m|)$ & Exact & PSLET & $(k,|m|)$ & Exact & PSLET \\ \hline
(1,7) & 20.3587 & 20.3587 & (1,7) & 23.5040 & 23.5040 \\ 
(0,9) & 20.3280 & 20.3280 & (0,9) & 23.2188 & 23.2188 \\ 
(3,2) & 18.5351 & 18.5351 & (3,2) & 23.0339 & 23.0339 \\ 
(2,4) & 18.4388 & 18.4388 & (2,4) & 22.2217 & 22.2217 \\ 
(1,6) & 18.3843 & 18.3843 & (3,1) & 21.8721 & 21.8715 \\ 
(0,8) & 18.3472 & 18.3472 & (1,6) & 21.7355 & 21.7355 \\ 
(3,1) & 16.6498 & 16.6498 & (0,8) & 21.3954 & 21.3954 \\ 
(2,3) & 16.4895 & 16.4895 & (3,0) & 21.3513 & 21.3140 \\ 
(1,5) & 16.4163 & 16.4163 & (2,3) & 20.6504 & 20.6504 \\ 
(0,7) & 16.3701 & 16.3701 & (1,5) & 20.0186 & 20.0186 \\ 
(3,0) & 14.9850 & 14.9881 & (0,7) & 19.6037 & 19.6037 \\ 
(2,2) & 14.5646 & 14.5646 & (2,2) & 19.2438 & 19.2438 \\ 
(1,4) & 14.4579 & 14.4579 & (1,4) & 18.3753 & 18.3753 \\ 
(0,6) & 14.3983 & 14.3983 & (2,1) & 18.1420 & 18.1420 \\ 
(2,1) & 12.6961 & 12.6961 & (0,6) & 17.8543 & 17.8543 \\ 
(1,3) & 12.5154 & 12.5154 & (2,0) & 17.6671 & 17.6660 \\ 
(0,5) & 12.4340 & 12.4340 & (1,3) & 16.8431 & 16.8431 \\ 
(2,0) & 11.0848 & 11.0883 & (0,5) & 16.1628 & 16.1628 \\ 
(1,2) & 10.6024 & 10.6024 & (1,2) & 15.4916 & 15.4916 \\ 
(0,4) & 10.4814 & 10.4814 & (0,4) & 14.5546 & 14.5547 \\ 
(1,1) & 8.7594 & 8.7594 & (1,1) & 14.4622 & 14.4622 \\ 
(0,3) & 8.5485 & 8.5485 & (1,0) & 14.0379 & 14.0381 \\ 
(1,0) & 7.2340 & 7.2362 & (0,3) & 13.0720 & 13.0720 \\ 
(0,2) & 6.6538 & 6.6538 & (0,2) & 11.7903 & 11.7903 \\ 
(0,1) & 4.8553 & 4.8553 & (0,1) & 10.8495 & 10.8496 \\ 
(0,0) & 3.4952 & 3.4968 & (0,0) & 10.4816 & 10.4816 \\ \hline\hline
\end{tabular}
\end{center}
\end{table}
\newpage 
\begin{table}[tbp]
\caption{ PSLET correlation energies (in $\hbar\protect\omega_o/2$ units)
for $k$=0, $|m|$=0,1,2,3,4,5 and $\protect\lambda$=0,1,2,4,6,8,10,12}
\begin{center}
\vspace{1cm} 
\begin{tabular}{|ccccc|}
\hline\hline
$|m|$ & $\lambda$=0 & $\lambda$=1 & $\lambda$=2 & $\lambda$=4 \\ \hline
0 & 2 & 3.4968 & 4.6391 & 6.4428 \\ 
1 & 4 & 4.8553 & 5.6557 & 7.1251 \\ 
2 & 6 & 6.6538 & 7.2872 & 8.4994 \\ 
3 & 8 & 8.5485 & 9.0864 & 10.1331 \\ 
4 & 10 & 10.4814 & 10.9564 & 11.8885 \\ 
5 & 12 & 12.4340 & 12.8638 & 13.7112 \\ \hline
& $\lambda$=6 & $\lambda$=8 & $\lambda$=10 & $\lambda$=12 \\ \hline
0 & 7.9373 & 9.2644 & 10.4816 & 11.6184 \\ 
1 & 8.4599 & 9.6938 & 10.8496 & 11.9425 \\ 
2 & 9.6480 & 10.7425 & 11.7903 & 12.7975 \\ 
3 & 11.1440 & 12.1226 & 13.0720 & 13.9947 \\ 
4 & 12.7978 & 13.6861 & 14.5547 & 15.4049 \\ 
5 & 14.5429 & 15.3599 & 16.1628 & 16.9525 \\ \hline\hline
\end{tabular}
\end{center}
\end{table}

\newpage 
\begin{table}[tbp]
\caption{ Same as table 3 for $k$=1.}
\begin{center}
\vspace{1cm} 
\begin{tabular}{|ccccc|}
\hline\hline
$|m|$ & $\lambda$=0 & $\lambda$=1 & $\lambda$=2 & $\lambda$=4 \\ \hline
0 & 6 & 7.2362 & 8.2945 & 10.0462 \\ 
1 & 8 & 8.7594 & 9.4879 & 10.8608 \\ 
2 & 10 & 10.6024 & 11.1913 & 12.3314 \\ 
3 & 12 & 12.5154 & 13.0233 & 14.0173 \\ 
4 & 14 & 14.4579 & 14.9110 & 15.8031 \\ \hline
& $\lambda$=6 & $\lambda$=8 & $\lambda$=10 & $\lambda$=12 \\ \hline
0 & 11.5189 & 12.8317 & 14.0381 & 15.1665 \\ 
1 & 12.1368 & 13.3327 & 14.4622 & 15.5362 \\ 
2 & 13.4252 & 14.4772 & 15.4916 & 16.4721 \\ 
3 & 14.9840 & 15.9254 & 16.8431 & 17.7388 \\ 
4 & 16.6773 & 17.5343 & 18.3753 & 19.2009 \\ \hline\hline
\end{tabular}
\end{center}
\end{table}
\newpage 
\begin{table}[tbp]
\caption{ Same as table 3 for $k$=2.}
\begin{center}
\vspace{1cm} 
\begin{tabular}{|ccccc|}
\hline\hline
$|m|$ & $\lambda$=0 & $\lambda$=1 & $\lambda$=2 & $\lambda$=4 \\ \hline
0 & 10 & 11.0883 & 12.0757 & 13.7327 \\ 
1 & 12 & 12.6961 & 13.3720 & 14.6650 \\ 
2 & 14 & 14.5646 & 15.1195 & 16.2015 \\ \hline
& $\lambda$=6 & $\lambda$=8 & $\lambda$=10 & $\lambda$=12 \\ \hline
0 & 15.1801 & 16.4736 & 17.6660 & 18.7833 \\ 
1 & 15.8856 & 17.0418 & 18.1420 & 19.1937 \\ 
2 & 17.2480 & 18.2613 & 19.2438 & 20.1978 \\ \hline\hline
\end{tabular}
\end{center}
\end{table}
\newpage

\begin{center}
{\bf {\Large Figures captions} }
\end{center}

{\bf Fig.1:} PSLET correlation energies for two interacting electrons in a
harmonic quantum dot vs $\lambda =\sqrt{2}l_{o}/a^{\ast }$ (the ratio of the
oscillator length $l_{o}$ and the effective Bohr radius $a^{\ast }$ . The
energies are normalized with the oscillator energy $\hbar \omega _{o}/2$.
The full lines represents states with $k=0$, dashed lines for $k=1$, and
dashed dotted lines for $k=2$.

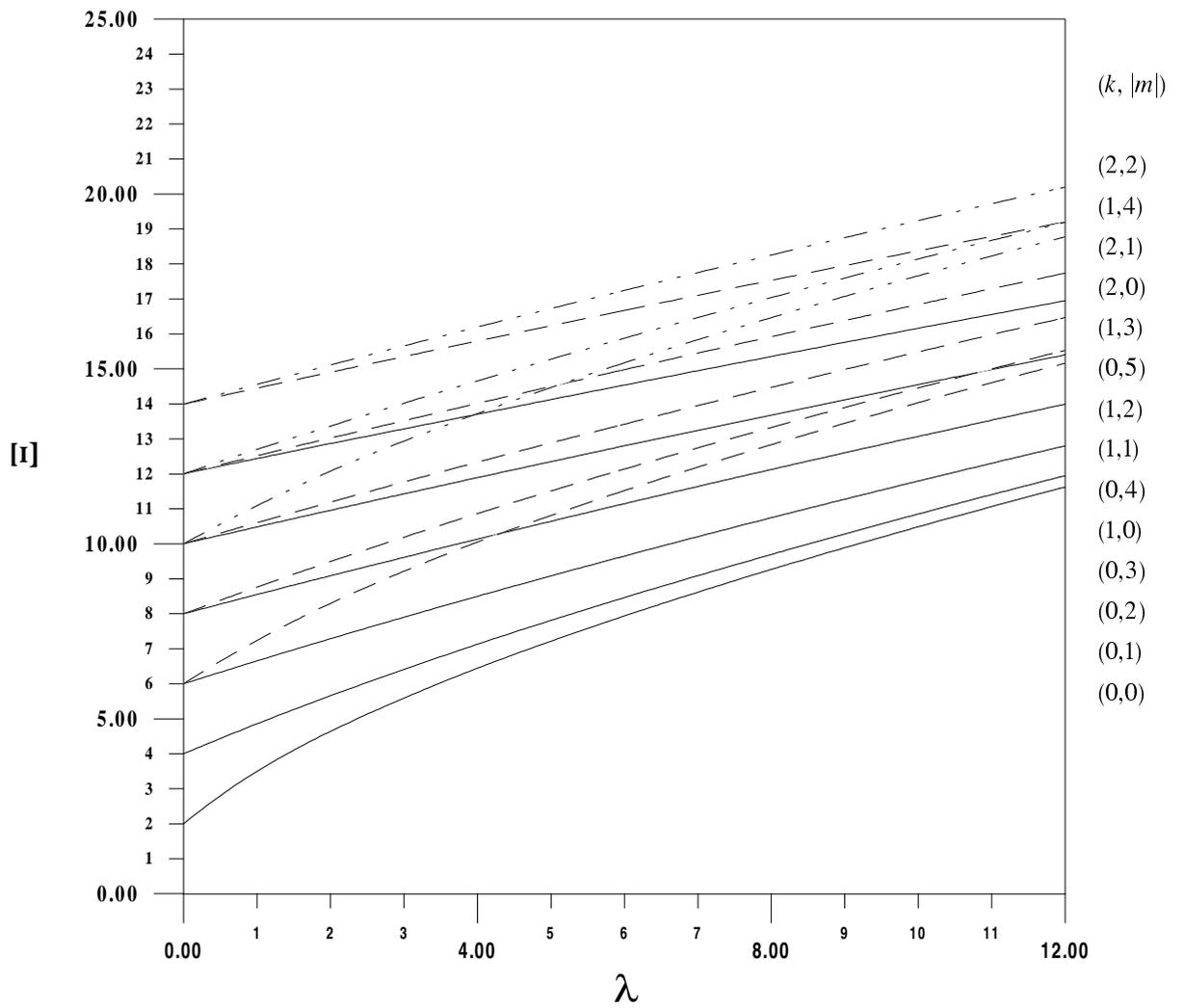
\end{document}